\documentclass[journal=jacsat,manuscript=article]{achemso}

\usepackage[version=3]{mhchem} 
\usepackage{xcolor}
\usepackage{comment}

\author{Leon Daniel}
\author{Dedi Sutarma}
\author{Osamah Kharsah}
\author{Charleen Lintz}
\author{Peter Kratzer}
\author{Marika Schleberger}
\email{marika.schleberger@uni-due.de}
\phone{
+49 203 379 1600/1601}
\fax{+49 203 379 2334}
\affiliation[1]{Faculty of Physics and CENIDE, University of Duisburg-Essen, Germany}

\title[PL of oleic-acid treated WS$2$]
{Mechanism of Oleic Acid-Mediated Sulfur Vacancy Healing in monolayer \ce{WS2}}


\begin{document}

\begin{tocentry}
\includegraphics[width=8 cm]{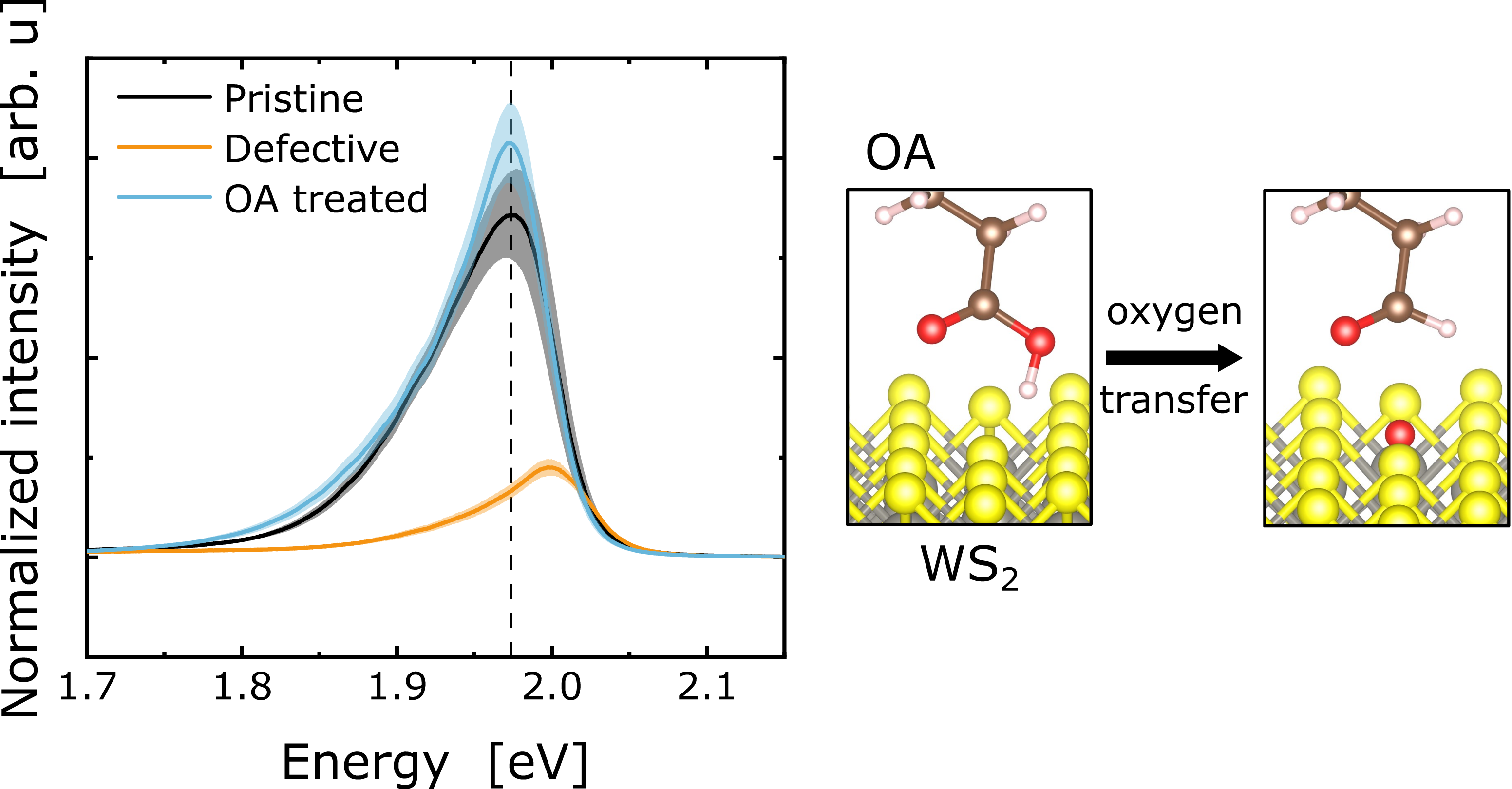}
\end{tocentry}

\begin{abstract}
We uncover the mechanism behind the enhancement of photoluminescence yield in monolayer \ce{WS2} through oleic acid treatment, a promising scalable strategy for defect healing. By inducing sulfur vacancies through thermal treatment and monitoring the changes in photoluminescence yield and emission spectra, we demonstrate that oleic acid heals the sulfur vacancy by providing substitutional oxygen. Using density functional theory calculations, 
we provide insight into the underlying mechanism governing the oleic acid-mediated sulfur vacancy healing process. Our findings suggest that effective defect passivation by oxygen doping can be achieved through chemical treatment, opening a pathway for oxygen doping in transition metal dichalcogenides. However, we also highlight the limitations of chemical treatment, which may only lead to small increases in photoluminescence yield beyond a certain point.
\end{abstract}

Transition metal dichalcogenides (TMDCs), particularly tungsten disulfide (\ce{WS2}), are promising candidates for next-generation optoelectronic and valleytronic devices due to their unique two-dimensional (2D) nature and exceptional optoelectronic properties \cite{Mak.2010, Gutierrez.2013}. As a direct bandgap semiconductor, monolayer \ce{WS2} has a bandgap of 2.4-2.7 eV \cite{Chernikov.2014, Gusakova.2017, Zhu.2015}, making it suitable for applications in the visible range \cite{Zhu.2015, Zeng.2013, Zhao.2013b}. Its high photoluminescence (PL) yield \cite{Xin.2022, Lee.2021, Kim.2021} enables a wide range of applications, from LEDs to photodetectors \cite{Andrzejewski.2020, Hutten.2021}. To develop \ce{WS2}-based applications, precise control and manipulation of defects, particularly vacancies, are essential. Vacancies can significantly affect the electronic structure of TMDCs and modify their optical and electronic properties, allowing for customization of device functionalities \cite{Bertolazzi.2017, Liu.2020, Sleziona.2024, Pollmann.2020, Bianchi.2024, Sleziona.2023, Pelella.2020}.

\begin{figure}[h]
\centering 
\includegraphics[width=1\textwidth]{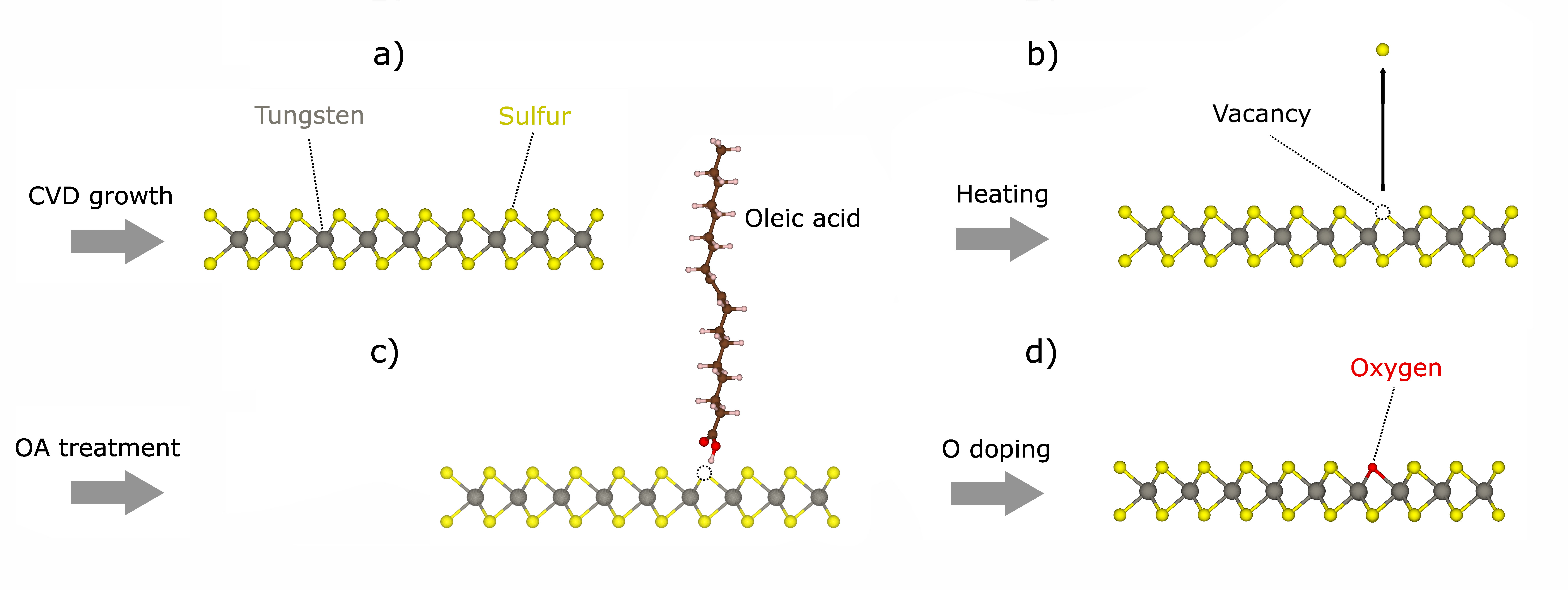}
    \caption{a) \ce{WS2} grown by CVD is heated to introduce single vacancies (b). The defective material is then treated with oleic acid (c), which substitutes the vacancy with an oxygen atom (d).}
\label{Figure0}
\end{figure}

Point defects in \ce{WS2}, such as sulfur vacancies,  introduce flat defect states in the bandgap, altering the material's native properties. These vacancies can act as a doping level or facilitate non-radiative recombination, affecting exciton recombination time and quantum efficiency (QE) \cite{Li.2016,Wang.2021,Cui.2021,Bianchi.2024}. To achieve specific material properties through defect engineering, both defect creation and passivation are crucial. Thermal treatment and ion irradiation have been used to induce vacancies in TMDCs, enabling precise control over defect type and density \cite{Hopster.2013,Madauss.2017,Schleberger.2018,Kozubek.2019,Mitterreiter.2021,Sleziona.2023,Asaithambi.2021}..

Vacancy healing, on the other hand, remains a significant challenge. Chemical treatments, such as super acid TFSI  \cite{Kiriya.2022,Amani.2015,Lu.2018}, have been explored as an effective and scalable approach to restore vacancies and enhance performance. However, handling these chemicals can be difficult due to their corrosive nature \cite{Cadore.2024}. A benign-by-design chemistry is therefore desirable. Oleic acid, a relatively weak acid and harmless alternative, has garnered significant interest due to its ability to effectively passivate vacancies in \ce{WS2} and enhance photoluminescence properties by increasing the QE of TMDCs, while preserving their n-doping characteristics \cite{Tanoh.2019, Tanoh.2021, Lin.2021, Wang.2022}.

However, the exact mechanism of the healing remains unclear. Our goal is to assess the efficacy of oleic acid in restoring the material's pristine characteristics and to reveal the underlying mechanism of the vacancy healing process. To this end, we used in-situ photoluminescence (PL) to analyze the evolution of vacancy density and band structure. By combining experiments on pristine and defective \ce{WS2} with first-principles calculations, we found that oleic acid molecules act as an oxygen source, saturating sulfur vacancies in \ce{WS2} and removing defect in-gap states, leading to improved PL (as schematically depicted in Figure~\ref{Figure0}). Our work thus demonstrates the feasibility and limits of using chemical treatments for defect management in \ce{WS2}, paving the way for optimized optoelectronic and valleytronic devices.

\begin{figure}[h]
\centering 
\includegraphics[width=0.7\textwidth]{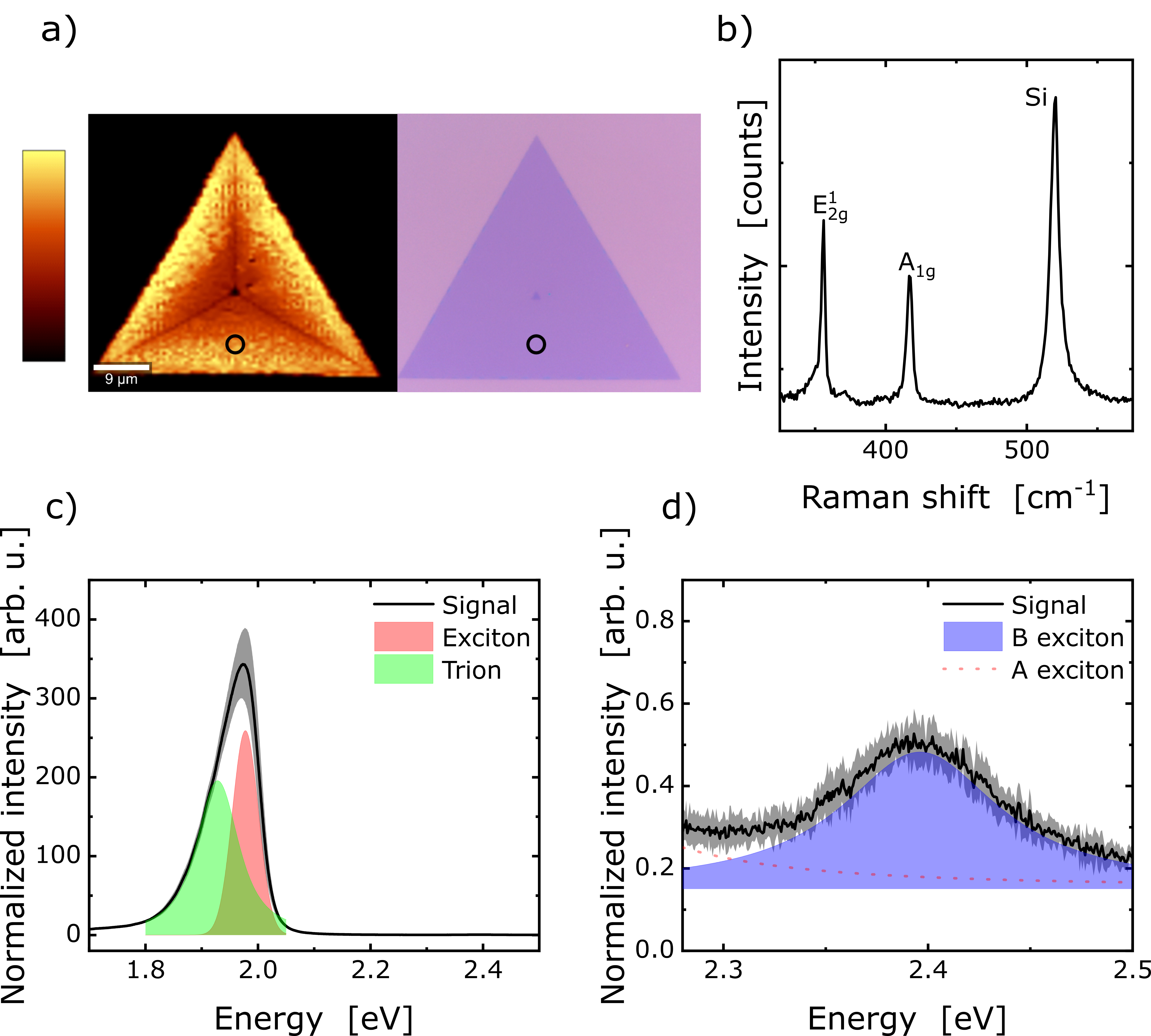}
    \caption[The system.]{a) Microscope image of a monolayer \ce{WS2} with corresponding PL map showing integrated intensity, with measurement locations marked. b) Raman spectra with assigned Raman modes. c) PL spectra of \ce{WS2} monolayer from 1.7 to 2.5~eV measured at 80~K, with fitted trion and exciton peaks. d) Zoomed-in section of the PL spectrum, highlighting the smaller B exciton signal.}
\label{Figure1}
\end{figure}

Our study begins with a characterization of pristine \ce{WS2} samples grown by   chemical vapor deposition (CVD, for details see Methods section). An optical microscopy image of a typical triangular monolayer flake is shown in Figure~\ref{Figure1}a) together with the integrated PL intensity map. Starting from the center, the PL intensity increases outwards, i.e.~along the direction of growth of the flake. This is indicative of strong oxygen doping \cite{Barja.2019,Luo.2022} as we use no hydrogen to bind excess oxygen. The substitution of vacancies with oxygen enhances PL emission by removing in-gap defect states, enabling non-radiative recombination \cite{Cui.2021, Wang.2021}. This is in agreement with DFT calculations showing that a vacancy saturated with an oxygen atom is highly favorable (see Supplementary Information and  \cite{Barja.2019, Cui.2021, Wang.2021, Hu.2019, Luo.2022}).

All further measurements were conducted at the spot with a strong PL intensity, marked with a black circle in Figure~\ref{Figure1}a). Figure~\ref{Figure1}b) displays a recorded Raman spectrum of the monolayer. By using a 457~nm laser, the 2LA~(\textit{M}) mode at 350.4~cm$^{-1}$ is strongly suppressed, enabling a clear view on the E$^{1}_{2g}$ ($\Gamma$) mode at 356.8~cm$^{-1}$ and A$_{1g}$ ($\Gamma$) mode at 416.9~cm$^{-1}$, confirming the monolayer \cite{Berkdemir.2013,Zhao.2013,Staiger.2015}. 

The pristine sample shows a strong emission at around 1.93~eV (Fig.~\ref{Figure1}c)), corresponding to the typical PL spectra for monolayer \ce{WS2} \cite{Gutierrez.2013,Zeng.2013, Zhao.2013b}. The PL signal is comparable to that of an exfoliated reference sample, indicating a low intrinsic defect concentration. The peak is asymmetric, suggesting the presence of trions \cite{Christopher.2017}.

The zoomed-in section of the PL spectra in Figure~\ref{Figure1}d) features the less intensive B exciton. This is consistent with the literature, where the A exciton emission is about 1,800 times stronger than the B exciton emission \cite{Zeng.2013,Zhao.2013b}. We modeled the band structure of a pristine \ce{WS2} monolayer via DFT calculations. Our calculations show a valence band splitting of 430 meV, which is 16 times higher than the conduction band splitting of 28 meV, consistent with literature values \cite{Zhu.2015, Zeng.2013, Zhao.2013b}. The predicted energy difference of the optical transition (402 meV) matches our experimental value of 420 meV ± 7 meV measured at 80 K very well.

\begin{figure}[h]
\centering 
\includegraphics[width=1\textwidth]{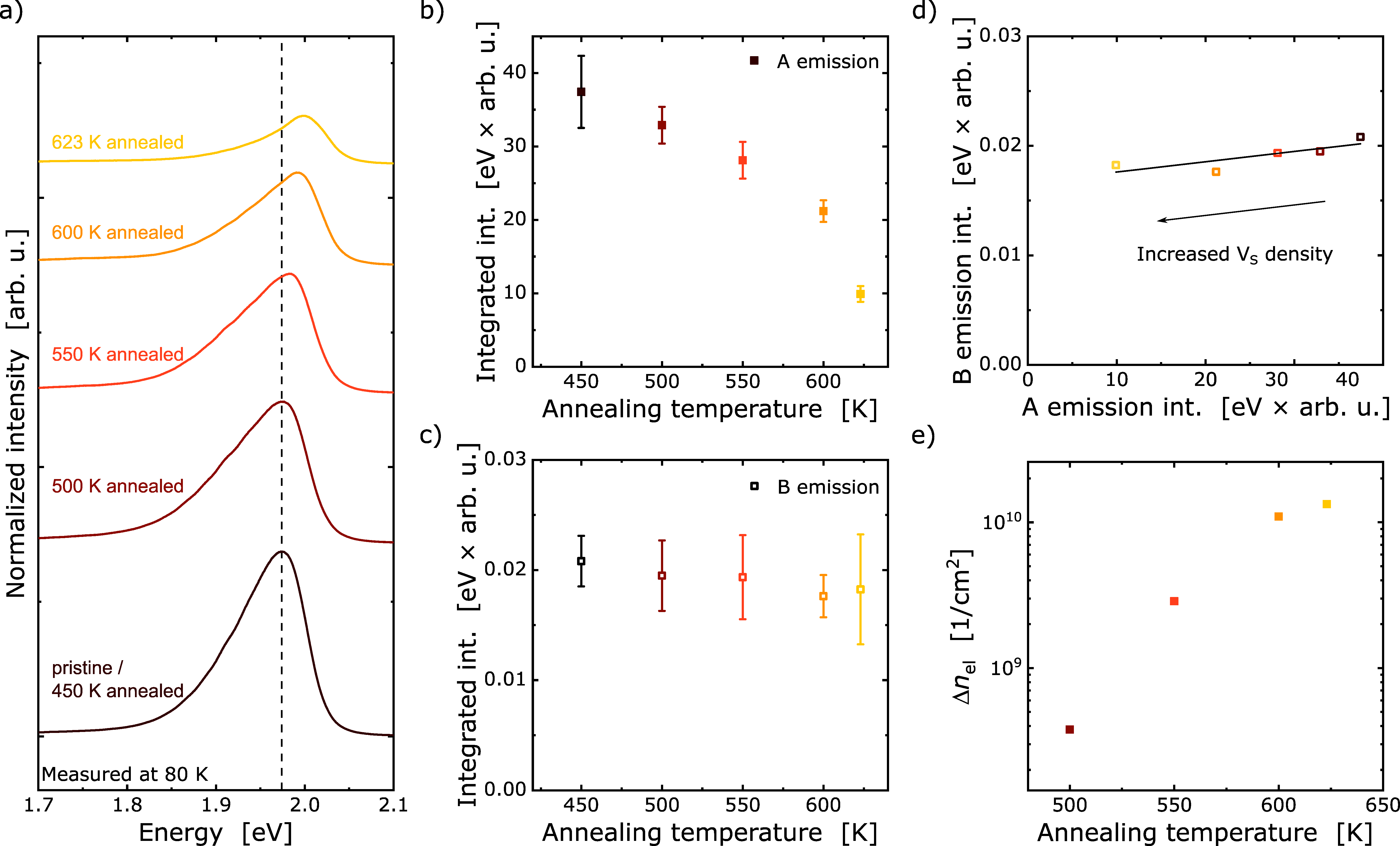}
    \caption[The system.]{a) Normalized PL spectra of \ce{WS2} for each heating step. b) and c) Signal intensity evolution of A and B emission with heating steps. d) B emission vs. A emission for the heated sample
    e) Evolution of free carrier density with increasing heating steps, calculated using the mass action model.}
\label{Figure2}
\end{figure} 
Next, we introduce sulfur vacancies by thermal processing, heating the sample to increasing temperatures and rapidly cooling to 80~K between each step (Methods section). In Figure~\ref{Figure2}a), the corresponding PL signal for the A-exciton is shown for selected temperature steps. The signal remains stable up to 450~K, consistent with other experiments \cite{Mitterreiter.2021}. However, with further temperature increase, the signal intensity decreases significantly, dropping to 26.5\% of the original intensity at 623~K. This is attributed to an increase in sulfur vacancies, which enable non-radiative recombination and reduce bright emission \cite{Li.2016, Wang.2021}. Supporting DFT calculations show that the formation energy for a sulfur vacancy is 2.72~eV, significantly lower than other types of vacancies. Additionally, the reaction enthalpy for sulfur removal in the presence of oxygen is -0.09~eV, indicating enhanced sulfur vacancy creation.

The B exciton, emitted by a different optical recombination path, shows no significant change in emission, as seen in Figure~\ref{Figure2}c). This suggests that the radiative recombination of the B exciton is unaffected by defects that influence non-radiative recombination of the A exciton. The ratio between A exciton emission and B exciton emission is often associated with defect density. For example, ion irradiation in MoS2 has been shown to decrease A exciton emission relative to B exciton emission \cite{Bertolazzi.2017}. McCreary et al. proposed a linear dependency of the intensity ratio of B and A emission in TMDCs:
$I(B) = b + a\times I(A)$ with $a$ and $b$ positive constants determined from  \ce{WS2} samples with various defect densities \cite{McCreary.2018}. we have plotted the intensity ratio of A~ and B exciton emission in Figure~\ref{Figure2}d). We determine $a = 0.00015 \pm 0.00005$, and from the intersection with the y-axis $b=0.015 \pm 0.002$.

To check if the correlation between A and B excitons is universal or defect-type-dependent, we compared the correlation in a sample irradiated with different fluences of 100~eV argon ions, which increases in-gap states \cite{Lu.2020}. The correlation differs significantly, and can be described by $I(B) = 0.0006 + 0.024 \times I(A)$, see SI. This suggests that individual defects can be differentiated based on the type of correlation between A and B excitons. Additionally, increasing sulfur vacancies should decrease the A exciton signal to almost zero while only slightly affecting the B exciton, indicating no large conversion from B excitons to A excitons.

With increasing vacancy density, the emission peak becomes increasingly asymmetric, attributed to increased trion emission at higher heating temperatures (see SI). This is consistent with the law-of-mass-action model, which describes the correlation between free charge carrier density and trion emission in TMDCs \cite{Ross.2013, Mouri.2013, Li.2017, Hichri.2017, Gaur.2019}.

\begin{equation}
\frac{N_{X}n_{el}}{N_{X^-}} = 
\frac{4m_{X}m_{e}}{\pi\hbar^2m_{X^-}} 
k_{B}T\exp{\left(-\frac{E_b}{k_{B}T}\right)}
\end{equation}

$N_{X}$ and $N_{X^-}$ are the concentration of free excitons and trions, $m_{e}$, $m_{X}$ and $m_{X^-}$ are the effective mass of the electron, free exciton and trion, $k_{B}$ and $T$ are the Boltzmann constant and temperature, and $E_b$ is the trion binding energy. 
We estimate the change in free charge carrier density $\Delta n_{el}$ with higher heating steps, as shown in Figure~\ref{Figure2}e), which suggests a maximum increase of up to $1.3\cdot 10^{10}$~cm$^{-2}$ for the final heating state. These carriers are attributed to ionized donor levels associated with S (or Se) vacancies \cite{Zhang.2021b, Shen.2022}.

\begin{figure}[h]
\centering 
\includegraphics[width=1\textwidth]{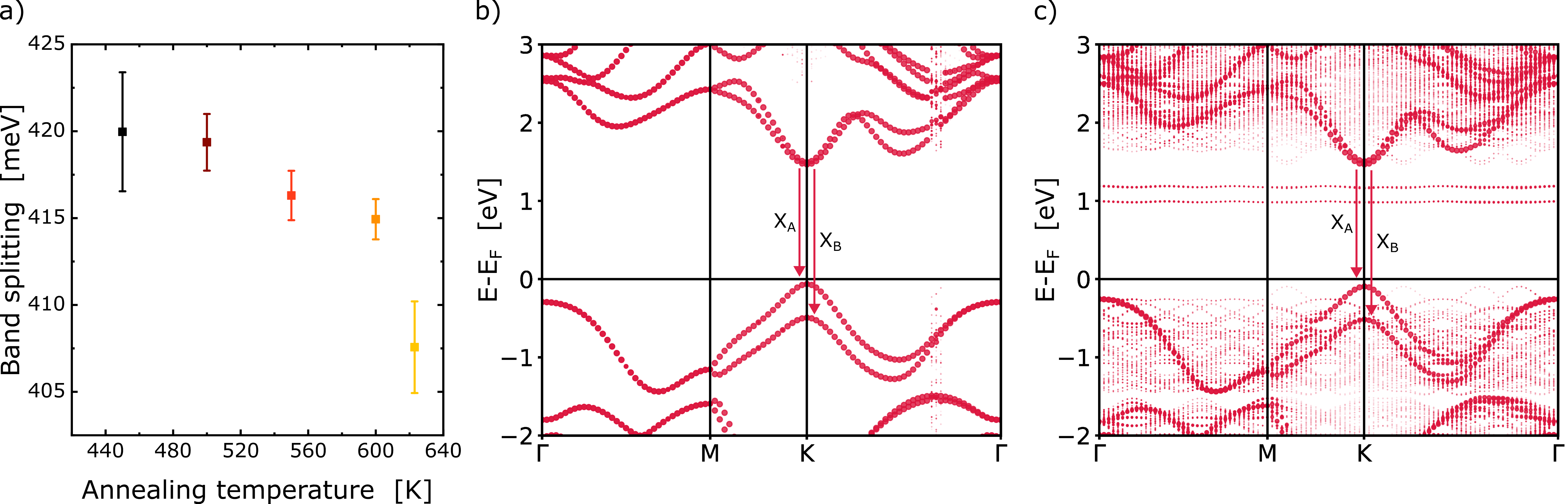}
    \caption[The system.]{a) Valence band splitting with increasing treatment temperature, with error bars representing one standard deviation. Band structure of \ce{WS2} in pristine condition (b) and with a V\textsubscript{S} vacancy (c). The energetic difference between A and B emission (X\textsubscript{A} and X\textsubscript{B}) decreases by 20.1~meV.}
\label{Figure3}
\end{figure} 

The signal position is analyzed to determine the splitting of conduction and valence bands. A significant shift in the A exciton maxima is observed with increasing defect density, as seen in Figure~\ref{Figure2}a), while the B exciton remains largely unaffected. The A-B-exciton splitting is plotted for different heating steps in Figure~\ref{Figure3}a). The origin of the band splitting is the spin-orbit coupling (SOC) induced mostly by tungsten \cite{Wang.2015, Rigosi.2016}. Sulfur vacancies have inconclusive effects on the band structure. However, some studies associate them with a red-shift due to higher trion emission \cite{Bianchi.2024} and DFT calculations suggest a slight increase in the band gap of \ce{WS2}~\cite{Wang.2021}.

We analyzed the electronic band structure of \ce{WS2} with various defects (V\textsubscript{S}, V\textsubscript{WS3}, O\textsubscript{S}). The results for pristine \ce{WS2} and V\textsubscript{S} are shown in Figure~\ref{Figure3}b) and \ref{Figure3}c). The vacancy induces two unoccupied states in the band gap and an occupied state in the valence band, leading to non-radiative recombination~\cite{Schuler.2019}. The calculations also suggest a decrease in the energetic difference between A and B emission of 20.1~meV, comparable to the experimental value. The removal of a WS\textsubscript{3} cluster or  oxygen substitution have only negligible effects, see SI.
\\

\begin{figure}[ht]
\centering 
\includegraphics[width=1\textwidth]{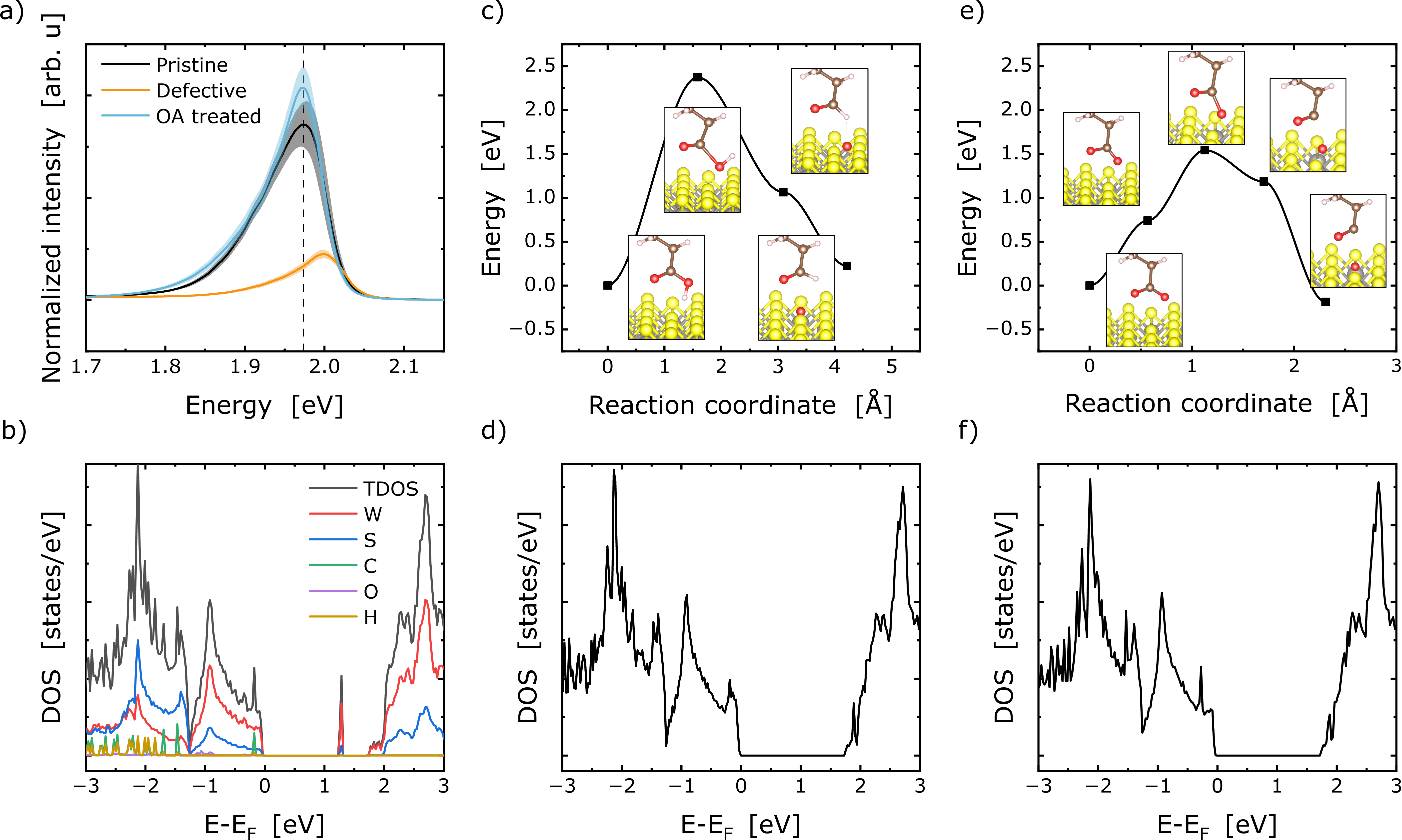}
    \caption[The system.]{a) Normalized PL signal at 80~K of a pristine CVD-grown sample after heating to 623~K and subsequent oleic acid treatment. b) Element-resolved density of states (DOS) of oleic acid adsorbed on a sulfur vacancy in \ce{WS2}. c) Reaction path of oleic acid with corresponding DOS (d) at the final reaction step. e) Reaction path of de-protonated oleic acid with corresponding DOS (f) at the final reaction step.}
\label{Figure4}
\end{figure} 

In the following, we examine the effect of oleic acid treatment on restoring the PL signal in defective samples. After thermal treatment, the PL intensity drops and shifts to the blue (Figure~\ref{Figure4}a). However, oleic acid treatment significantly increases the PL intensity, consistent with previous reports \cite{Tanoh.2019, Tanoh.2021, Lin.2021, Wang.2022}. The A exciton emission increases by a factor of 4 on average, while the B exciton emission increases by 19\%. This is attributed to the elimination of defect-induced in-gap states.
The recovered PL emission of the A exciton exceeds the pristine sample's by 8\%. This might be due to a step in the oleic acid treatment, which involves washing in toluene. This might reduce contaminations present on \ce{WS2}, enhancing electron-hole recombination. We also observe a slight reduction in doping concentration by up to $1.12 \times 10^{10}$~cm$^{-2}$ and a shift in A emission spectra back to its original position (B emission remains unaffected).

We found that oleic acid treatment of pristine \ce{WS2} samples did not significantly improve PL emission, unlike previous reports. We attribute this to oxygen atoms occupying sulfur vacancies, removing in-gap states and preventing oleic acid from binding to or reacting with them. This suggests that oleic acid's presence alone does not enhance PL emission, contradicting a purely physical interaction mechanism as, e.g.,~charge transfer resulting in reduced trion formation. We suggest that PL recovery is due to oxygen atoms provided indirectly during oleic acid treatment, which saturate V\textsubscript{S} sites. This mechanism is supported by the fact that only point defects, not larger defects (see SI), can be healed.

While O\textsubscript{2} molecules might be present on the surface, they do not dissociate readily at V\textsubscript{S} \cite{Bianchi.2024.O2, Luo.2022}. Therefore, another source for oxygen must be involved. To support our claim, that the oleic acid provides the oxygen, we performed DFT calculations on the reduction of oleic acid to a fatty aldehyde, releasing an oxygen atom. Figures \ref{Figure4}b) and \ref{Figure4}c) display the calculated minimum energy path for the reduction of oleic acid and deprotonated oleic acid at V\textsubscript{S}, respectively. While the former reaction is found to be endothermic by 0.22~eV, the latter is exothermic by 0.19~eV. As shown in Figure~\ref{Figure4}b), oleic acid reduction proceeds by elongation of the bond between the hydroxy group and the acyl head, with an energy barrier of 2.37~eV. The reduction of deprotonated oleic acid is more likely, see Figure~\ref{Figure4}c): After breaking the carbon-oxygen bond, the oxygen atom inserts itself into the sulfur vacancy, with an energy barrier of 1.54~eV. We note that thermal activation suffices to overcome the barrier of 1.54~eV even at room temperature. Eventually, we verified that for both reactions, the in-gap states associated with V\textsubscript{S} are indeed removed. This is concluded from Figures \ref{Figure4}e) and f), showing the calculated density of states in the final state of the reactions.

Supplementary DFT calculations have been carried out to address the transfer of a hydrogen atom from oleic acid to the sulfur vacancy. These studies are relevant since alternative passivation strategies using super-acids \cite{Yamada.2020} such as TFMS \cite{Feng.2021} or TFSI \cite{Kiriya.2022, Amani.2015, Lu.2018} point to the role of hydrogen transfer for enhancing PL yield. Here we note that oleic acid is known to be a weak acid (pKa~9.85 \cite{Kanicky.2002}), making hydrogen transfer {\it a priori} less likely in this case. Thus, different mechanisms are likely to be at work in super-acids as compared to oleic acid. In line with this observation, our own DFT calculations (see SI) show that oleic acid, both in its integral or deprotonated form, can adsorb either on-top of or inside the sulfur vacancy, but in all four geometries studied by us, the passivation of the dangling bonds was incomplete. However, deprotonation at V\textsubscript{S} could still play a role as a first step followed by reduction to the fatty aldehyde, as described above. We note that other chemicals, in particular sulfur-containing ones \cite{Roy.2018}, e.g.~thiols \cite{Schwarz.2023}, may provide an alternative route to PL recovery, as has been discussed in the literature.

In summary, we systematically investigated the impact of thermally induced vacancies on A and B emission through \textit{in-situ} measurements. We observed a strong effect on A excitons, while B excitons were almost unaffected by vacancy-induced defect states, suggesting that the correlation between the two excitonic signals is defect-type dependent. The energetic difference between A and B emission is influenced by the defect density. Our results demonstrate oleic acid's effectiveness in restoring the material's optoelectronic properties. We showed that oleic acid primarily saturates existing sulfur vacancies. Our calculations indicate that the PL enhancement observed with oleic acid is primarily due to the introduction of substitutional oxygen, which heals vacancies and removes in-gap states. No further significant improvement beyond oxygen substitution could be achieved.

\section{Methods}
\subsection{Sample Preparation}
\ce{WS2} samples were grown via chemical vapor deposition (CVD) on SiO$_2$/Si substrates. To prepare the growth substrates, an aqueous solution was mixed from 3.2 mL of ammonium metatungstate (AMT, Sigma Aldrich), 2 mL of OptiPrep (Sigma Aldrich), and 0.8 mL of DI water. This solution was spin-coated onto p-doped Si substrates with a 285 nm SiO$_2$ layer \cite{An.2022}, followed by spin-coating cholic acid sodium salt (Sigma Aldrich) and subsequent annealing at 500°C for 45 minutes to convert AMT to WO$_3$.

The CVD growth was performed in a three-zone tube furnace with a quartz tube (ThermConcept) connected to an Ar source and exhaust, providing a flow rate of 500 sccm. The first zone was heated to 170~$^\circ$C with 350~mg of sulfur (S powder, Sigma Aldrich, 99.98\%), while the growth substrates were positioned in the second zone at 725~$^\circ$C for 60 minutes to form \ce{WS2}. The third zone was kept at 600~$^\circ$C to remove reaction by-products.

Using transfer-free samples eliminates defects formed during transfer processes, such as localized strain caused by impurities on intermediate surfaces \cite{Lee.2020}.

\subsection{Thermal Treatment}
The sample was placed in a Linkam THMS350V vacuum chamber, achieving a medium vacuum of $5\cdot 10^{-3}$~mbar. It was heated to various temperatures (up to 623 K) for 30 minutes, then rapidly cooled to 80~K using liquid nitrogen. PL measurements were taken at 80~K once equilibrium was reached.

\subsection{PL Measurements}
PL measurements were performed using a Witec Alpha300 R setup. Spectra were averaged from 10 measurements with a 457 nm laser, focused to $\approx$~830 nm spatial resolution. The full spectrum (1.5-2.7 eV) was taken to normalize the signal after background subtraction, using the Si substrate's secondary Raman mode (985~cm$^{-1}$). A constant laser power of 1.25$\times$10$^5$~W/cm$^2$ was used, with the laser spot readjusted and refocused for each measurement to minimize signal fluctuations. All measurements were conducted at 80~K under vacuum conditions.

\subsection{Oleic Acid Post-Treatment}
We followed Tanoh et al.'s instructions \cite{Tanoh.2019} for oleic acid treatment. After heating, the sample was transferred to a nitrogen-filled glovebox and placed on a heating plate kept at 25~$^\circ$C. Degassed oleic acid was carefully dropped onto the sample, covering the substrate. After 16 hours, the oleic acid was washed away with toluene and dried with nitrogen.

\subsection{DFT Calculations}
First-principles calculations were performed using Density Functional Theory (DFT) with VASP 6.3.0 version package \cite{Kresse.1996, Kresse.1999}. The interactions  between ions and valence electrons were described using the projector-augmented wave (PAW) \cite{Blochl.1994} method, and electronic exchange and correlation were treated by the generalized gradient approximation of Perdew, Burke, and Ernzerhof (PBE)\cite{Perdew.1996}. We employed a plane wave cut-off energy of 500~eV with an additional correction to the van der Waals dispersion interaction in the form of Becke-Johnson damping function D3\cite{Grimme.2011}. The material models including the defected system were built in a $6\times 6\times 1$ supercell with 15 \AA\:vacuum to avoid interaction with the periodic images, which corresponds to a defect density of $2.1\times 10^{13}$~cm$^{-2}$. A $9\times  9\times 1$ Monkhorst-Pack k-point grid was used to sample the 2D Brillouin zone. The unfolding of the  band structure was computed using the VASPKIT code \cite{Wang.2021b}. We constructed the minimum energy path of our reactions using the Nudged Elastic Band (NEB) method \cite{Henkelman.2000}.

\subsection{Large Language Models}
We used  an LLM (Meta Llama 3.1 8B Instruct) to improve the linguistic quality of the manuscript. Neither this nor any other LLM was used to generate scientific content.

\section{Author Contributions}

M.S.~and P.K.~ conceived and supervised the project. L.D. designed the experimental research strategy and performed the measurements. D.S.~conducted the theoretical calculations. O.K. fabricated the samples. O.K. and C.L. assisted with performing the measurements.
 The manuscript was written through contributions of all authors. All authors have given approval to the final version of the manuscript.
\\
\\
\noindent
Notes\\
\noindent
The authors declare no competing ﬁnancial interest.

\begin{acknowledgement}
This work was funded by the Deutsche Forschungsgemeinschaft (DFG, German Research Foundation) - project numbers 461605777 (IRTG 2803 2D MATURE) and 429784087. The authors acknowledge support by technical staff, especially Anke Hierzenberger, and helpful discussions with German Sciaini. The authors gratefully acknowledge the computing time granted by the Center for Computational Sciences and Simulation (CCSS) of the University of Duisburg-Essen and provided on the supercomputer magnitUDE (DFG Grant No. INST 20876/209-1 FUGG and INST 20876/243-1 FUGG) at the Zentrum f{\"u}r Informations und Mediendienste (ZIM). 
\end{acknowledgement}

\bibliography{bibliography}

\end{document}